\documentclass[12pt]{article}
\usepackage{graphicx,amsmath,amssymb}
\usepackage{indentfirst}
\usepackage[usenames]{color}
\usepackage[colorlinks=true, urlcolor=navyblue, linkcolor=navyblue, citecolor=navyblue]{hyperref}
\usepackage{epstopdf}
\usepackage{enumerate}
\usepackage{appendix}
\usepackage{lineno}
\usepackage{setspace}

\definecolor{navyblue}{rgb}{0,0.08,0.45}
\definecolor{darkred}{rgb}{0.7,0.0,0.0}
\definecolor{darkgreen}{rgb}{0,0.6,0.2}

\newcommand{\beq}{\begin{equation}}
\newcommand{\enq}{\end{equation}}
\newcommand{\beqa}{\begin{eqnarray}}
\newcommand{\beqast}{\begin{eqnarray*}}
\newcommand{\enqa}{\end{eqnarray}}
\newcommand{\enqast}{\end{eqnarray*}}
\newcommand{\nn}{\nonumber}

\newcommand{\bec}{\begin{center}}
\newcommand{\enc}{\end{center}}
\newcommand{\beqo}{\begin{quote}}
\newcommand{\enqo}{\end{quote}}
\newcommand{\bem}{\begin{minipage}}
\newcommand{\enm}{\end{minipage}}

\newcommand{\req}[1]{(\ref{#1})}

\newcommand{\half}{\textstyle \frac{1}{2}}

\newcommand{\la}{\lambda}

\newcommand{\si}{\sigma}

\newcommand{\vp}{\varphi}

\newcommand{\La}{\Lambda}

\begin{document}

\vspace{15pt}

\begin{center}
{\huge  Hadron Physics from Superconformal}

\vspace{10pt}

{\huge    Quantum Mechanics}

\vspace{10pt}

{\huge    and its Light-Front Holographic Embedding}

\end{center}

\vspace{15pt}

\centerline{Guy F. de T\'eramond}

\vspace{3pt}

\centerline {\it Universidad de Costa Rica, San Jos\'e, Costa Rica~\footnote{{Invited talk presented at Light-Cone 2015: Theory and experiment for hadrons on the light front, September 21-25, 2015, INFN, Frascati, Italy\\
\href{mailto:gdt@asterix.crnet.cr}{\tt
\hspace{12pt} gdt@asterix.crnet.cr}}}}

%\centerline{\today}

\vspace{20pt}

\begin{abstract}

The complex nonperturbative color-confining dynamics of QCD  is well captured in a semiclassical effective theory based on superconformal quantum mechanics and its extension to the light-front.  I  describe here how this new approach to hadron physics incorporates confinement, the appearance of nearly massless pseudoscalar particles, and Regge spectroscopy consistent with experiment. It  also gives remarkable connections between the meson and baryon spectrum across the light and heavy-light hadron spectrum. I also briefly discuss how higher spin states are consistently described in this framework by the holographic embedding of the superconformal theory in a higher dimensional semiclassical gravity theory.

\end{abstract}

\newpage

\tableofcontents

\section{Introduction}

Quantum Chromodynamics (QCD) describes the interactions of quark and gluons with remarkable  success in the high energy regime where asymptotic freedom allows accurate computations using perturbation theory. However, in the low energy domain, where hadrons are the relevant degrees of freedom, QCD is nonperturbative and understanding the mechanism of confinement is still an unsolved problem. In fact, no analytic solution to the problem of confinement in QCD has been found yet in the physical realm~\cite{Brambilla:2014jmp}.  According to the Kinoshita-Lee-Nauenberg theorem,  perturbative QCD cannot describe  confinement at any order in perturbation theory,  at least in a simple way~\cite{Kinoshita:1962ur, Lee:1964is}. Nonperturbative methods such as lattice field theory, effective field theories, Schwinger-Dyson equations or the gauge/gravity correspondence, are thus needed to study confinement dynamics~\cite{Brambilla:2014jmp, Greensite:2011zz}.

The problem of confinement is vastly complex. The QCD Lagrangian in the limit of massless quarks has no scale; still confinement and a mass gap should emerge from the quantum theory built upon the classical QCD conformal theory.  The increase of the QCD coupling in the infrared regime implies that an infinite number of quark and gluons are dynamically intertwined, and a description of the strong dynamics in terms of the fundamental fields appearing in the QCD Lagrangian becomes analytically intractable. Indeed, the dynamical problem may become undecidable:  In QCD this means that it is not possible to know whether the system is truly gapless, or whether increasing the lattice size would reveal it to be gapped~\cite{Cubitt:2015nat}.

Holographic methods provide new tools for the study of strongly correlated quantum systems. The AdS/CFT correspondence between gravity on a five-dimensional anti-de Sitter (AdS) space and conformal field theories (CFT) in physical space-time~\cite{Maldacena:1997re} is an explicit realization of the holographic principle. The gauge/gravity correspondence leads to a semiclassical gravity approximation for a strongly-coupled non-abelian quantum field theory, thus providing new physical insights into its nonperturbative dynamics.

The holographic mapping to light-front (LF) physics is particularly useful since quantization in the front form of dynamics~\cite{Dirac:1949cp} provides a relativistic and frame independent Hamiltonian framework to describe bound-state hadronic systems~\cite{Brodsky:1997de}. In fact, since the structure of the LF vacuum is particularly simple, it is possible to retain a probabilistic quantum-mechanical interpretation in terms of hadronic wave functions.   One can show that the eigenvalue equations in AdS space have a precise mapping to the relativistic  semiclassical bound-state equations in the light front~\cite{deTeramond:2008ht}.  This connection also gives an exact relation between the holographic variable $z$ of AdS space and the invariant impact separation light-front variable $\zeta$~\footnote{Light-front holographic methods were introduced by matching the electromagnetic form factor in AdS space with the corresponding light-front expression in physical space-time~\cite{Brodsky:2006uqa}. The invariant distance squared for a two-particle bound state is $\zeta^2 = x(1-x) b^2_\perp$, where $x$ is the constituent momentum fraction, and $b_\perp$ is the transverse  separation between the constituents.}.

The holographic mapping implies that the non-trivial geometry of AdS space encodes all the kinematical aspects, whereas the modification of the action in AdS space --described in terms of a dilaton profile $\varphi(z)$, includes the confinement dynamics and determines the effective LF potential $U(\zeta)$ in the light-front bound-state equations~\cite{deTeramond:2013it}.  The embedding of LF wave equations in AdS leads to an extension of the LF potential $U$ to arbitrary spin from conformal symmetry breaking in the AdS action~\cite{deTeramond:2013it, Gutsche:2011vb}~\footnote{The potential $U$ is related to the dilaton profile by~\cite{deTeramond:2013it, deTeramond:2010ge}
$ U(\zeta,J)=\frac{1}{2} \vp''(\zeta) + \frac{1}{4} \vp'(\zeta)^2 +\frac{2 J-3}{2 \zeta}\,\vp'(\zeta)$, where $J$ is the total angular momentum of the hadron.}.
The light-front effective potential $U$ acts on the valence state and incorporates and infinite number of higher Fock states and retarded interactions~\cite{Pauli:1998tf}, but its derivation from QCD represents an unsurmountable task.

Since the light-front  semiclassical approximation to the QCD Hamiltonian equation leads to a one-dimensional quantum field theory (QFT), it is natural to ask if there is a compelling connection of conformal quantum mechanics (QM)~\cite{deAlfaro:1976je} with hadronic physics.  Indeed, Brodsky, Dosch and I have shown recently that  one can introduce a gap and a confinement scale $\sqrt \la$ by extending conformal QM to the light front~\cite{Brodsky:2013ar}.   The precise mapping of conformal QM to the light-front bound-state equations uniquely determines the form of the LF confinement potential $U$ --and thus the dilaton profile $\varphi$, while keeping  the conformal invariance of the action preserved~\cite{deAlfaro:1976je}.  This formalism uniquely determines the confinement potential $U$ to have the form of a harmonic oscillator in the light-front transverse plane~\footnote{For light-quark masses the harmonic oscillator potential in the light front is equivalent  to the usual  linear confining potential at large separation distances in the instant form of dynamics~\cite{Trawinski:2014msa}.}. The resulting model is very successful in describing a wealth of hadronic data~\cite{Brodsky:2014yha}.

%%%%%%%%%%%%%%%%
\begin{figure}[ht]
\begin{center}
\includegraphics[width=8.6cm]{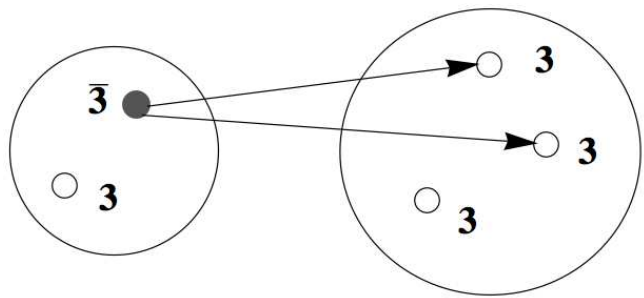}
\end{center}
\caption{\label{MNSUSY} \small In QCD hadronic dynamical supersymmetry is rooted in the dynamics of color $SU(3)$ where $\bf \bar 3 \sim \bf 3 \times \bf 3$.}
\end{figure}
%%%%%%%%%%%%%%%%%%

More recently Dosch, Brodsky and I have shown how the extremely complex nonperturbative QCD dynamics is well captured in a semiclassical effective theory based on superconformal quantum mechanics~\cite{Akulov:1984uh, Fubini:1984hf} and its extension to light-front physics~\cite{deTeramond:2014asa, Dosch:2015nwa, Dosch:2015bca}. This new approach to hadron physics incorporates two basic aspects one expects from QCD, but whose understanding has remained elusive: confinement and the appearance of nearly massless pseudoscalar particles. Furthermore, this framework gives remarkable connections between the light meson and nucleon spectra as well as predictions across the full heavy-light hadron spectra, where heavy quark masses break the conformal invariance, but the underlying supersymmetry  still holds. It is important to recall that  in the context of hadronic physics the supersymmetric relations  are not a consequence of supersymmetric QCD at the level of fundamental fields  but an emergent dynamical supersymmetry from color $SU(3)_C$. This relies on the fact that in $SU(3)_C$ a diquark can be in the same color representation as an antiquark, namely a $\bf \bar 3 \sim \bf 3 \times \bf 3$~\cite{Catto:1984wi} as illustrated in Fig.  \ref{MNSUSY}.  Hadronic supersymmetry was introduced by Miyazawa~\cite{Miyazawa:1966mfa}.

\section{Superconformal quantum mechanics and light-front bound-state equations}

We construct  a superconformal algebraic structure in one dimension by introducing the fermionic operators  $Q$,
$Q^\dagger$, $S$ {and} $S^\dagger$~\cite{Fubini:1984hf}
\beqa
\half\{Q,Q^\dagger\} & \! \! = \! \! & H,   \hspace{80pt}  \half\{S,S^\dagger\} =K, \hspace{20pt} \nn \\
\{Q,S^\dagger\} & \! \! =  \! \! & f  - B + 2 i D,  \hspace{30pt}  \{Q^\dagger,S\}  = f  - B - 2 i D,
\enqa
where  $Q$ is represented by 
$$Q =  \psi^\dag  \left(- \frac{d}{dx} + \frac{f}{x} \right)   \quad  {\rm and}  \quad   Q^\dagger =  \psi \left( \frac{d}{dx} + \frac{f}{x} \right),$$ 
with  $f$ a dimensionless quantity. The generator $S$ is given by $S=\psi^\dag x$ and $S^\dag=\psi x$, while $B=\frac{1}{2}[\psi^\dag,\psi]$ is a bosonic operator with $\{\psi,\psi^\dag\}=1$.  The three generators of translation, dilatation and the special conformal transformation $H$, $D$ and $K$
\beqa
 H &\!  =  &\!  \frac{1}{2} \left( - \frac{d^2}{d x^2}  + \frac{f^2 + 2 B f}{x^2} \right),   \nn \\
 D &\!  =   &\!  \frac{i}{4} \left(\frac{d}{dx} x + x \frac{d}{d x} \right), \nn \\
 K &\!  =   &\!  \half x^2, 
\enqa
 satisfy the conformal algebra:
\beq
[H,D]= iH, \qquad [H,K]= 2 i D, \qquad  [K,D]=-i K . 
\enq

Following Fubini and Rabinovici~\cite{Fubini:1984hf} we define a new fermionic operator $R$, a linear combination of the generators $Q$ and $S$,
\beq
R_\la = Q + \la S,    \hspace{30pt} R_\la^\dagger = Q^\dagger + \la S^\dagger,
\enq
which generates a new  Hamiltonian $G$
\beq \label{G}
G =\{Q_\la,Q_\la^\dagger\},
\enq
where by construction
\beq
\{R_\la,R_\la\} =   \{R_\la^\dagger, R_\la^\dagger\} = 0 \hspace{20pt}  {\rm and} \hspace{20pt}   [R_\la, G]  = [R_\la^\dagger, G] = 0,
\enq
a set of relations which closes a graded algebra $sl(1/1)$, identical to Witten's supersymmetric quantum mechanics~\cite{Witten:1981nf}. Indeed, since the Hamiltonian $G$ commutes with $R_\la$; it thus follows that the states  $\vert \phi \rangle$ and $R^\dagger  \vert \phi \rangle$ have identical eigenvalues: If $ \vert \phi_E\rangle$ is an eigenstate of $G$ with $E\neq 0$,  $G\,|\phi_E\rangle = E \,|\phi_E\rangle$, then $G \,R^\dagger_\la  \,  |\phi_E\rangle = R^\dagger_\la \,G\, |\phi_E\rangle=E\,  R^\dagger_\la |\phi_E\rangle$,  and thus  $R^\dagger_\la  \,  |\phi_E\rangle$ is also an eigenstate  of $G$ with the same eigenvalue.  Since the dimensions of the generators $Q$ and $S$ are different, a scale $\la$ is introduced in the Hamiltonian in analogy with the previous treatment by de Alfaro, Fubini and Furlan~\cite{deAlfaro:1976je}.

In a Pauli matrix representation Eq. \req{G} is given by 
$$G = 2 H + 2 \la^2 K + 2 \la\left(f - \si_3 \right),$$ 
leading to the eigenvalue equations~\cite{Dosch:2015nwa}
\beq  \label{phi1}
\left(- \frac{d^2}{d x^2} + \la ^2 \,x^2+ 2 \la \, f  - \la + \frac{4 (f+\half)^2 - 1}{4 x^2}\right) \phi_1 = E  \, \phi_1,
\enq
\beq  \label{phi2}
 \left(- \frac{d^2}{d x^2} + \la^2 \, x^2 + 2
\la \, f  + \la  + \frac{4 (f-\half)^2 -1}{4 x^2}\right) \phi_2 = E  \, \phi_2,
\enq
with identical eigenvalues for $f \ge \half$ and $\la > 0$:  
$$E_n =  4 \la \left( n + f + 1/2 \right).$$

We now compare Eqs. \req{phi1} and \req{phi2} with the leading twist LF holographic equations which describe the pseudoscalar and nucleon orbital and radial excitation spectrum~\cite{Brodsky:2014yha}
\beq \label{phiM}
 \left(-\frac{d^2}{d\zeta^2} + \frac{4 L_M^2 -1}{4 \zeta^2} + \la_M^2\, \zeta^2 + 2 \la_M (L_M - 1)   \right)\phi_{Meson} =  M^2 \, \phi_{Meson},
\enq
\beq  \label{phiN}
\left(-\frac{d^2}{d\zeta^2} + \, \frac{4 L_N^2 -1}{4 \zeta^2} + \, \la_N^2\, \zeta^2 + \,  2 \la_N (L_N +1)   \right)\phi_{Nucleon} =  M^2 \, \phi_{Nucleon} ,
\enq
with eigenvalues
\beqa
M^2_M & \! = \! & 4 \la_M \left(n + L_M \right), \label{M2M}\\
M^2_N  & \! = \! & 4 \la_N \left(n + L_N + 1\right). \label{M2N}
\enqa

Since $R$ is a fermion operator  which relates  the upper and lower components in the multiplet  
$\vert\phi \rangle = \rm \left(\begin{array}{c} \phi_1\\ \phi_2\end{array} \right),$
\beq R_\la   =\left(\begin{array}{cc}
0&-\frac{d}{dx}+ \frac{f}{x} + \la\,x\\
0&0
\end{array}\right),
\qquad 
 R_\la^\dagger=\left(\begin{array}{cc}
0&0 
\\
\frac{d}{dx}+ \frac{f}{x} + \la\,x&0
\end{array}\right),
\enq
it is tempting to relate the meson and nucleon bound-state wave functions by the supercharges within the superconformal algebra. Identifying the upper component equation \req{phi1} with the  meson LF bound-state equation \req{phiM}  and the lower component equation \req{phi2}  with the nucleon equation \req{phiN} we find, upon the substitution $x \to \zeta$ and $E \to M^2$ the conditions $\la = \la_M = \la_N$,  and  $f = L_N + \half = L_M - \half $. Thus the remarkable relation $L_M = L_N + 1$ between the orbital angular momentum of the meson and nucleon superpartners~\cite{Dosch:2015nwa}. This result is in agreement with the  light-front holographic QCD spectral formulas Eqs. \req{M2M} and \req{M2N} for pseudoscalar mesons and nucleons respectively.  The superconformal approach has, however, the advantage that mesons and nucleons are treated on the same footing and additional constant terms in the confinement potential are determined from the onset~\footnote{In Ref.~\cite{deTeramond:2014asa}  superconformal quantum mechanics was used to describe the nucleon spectrum. In this case, the supercharges  relate the positive and negative chirality components of the nucleon wave functions.}.

%%%%%%%%%%%%%%%%
\begin{figure}[h]
\begin{center}
\includegraphics[width=7.6cm]{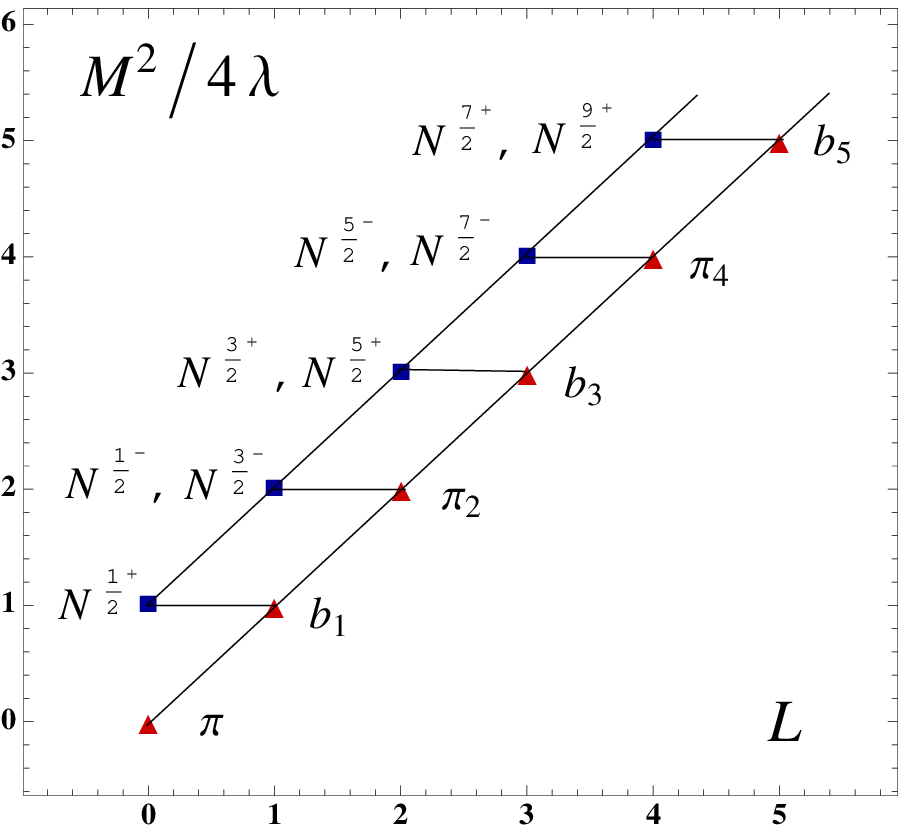}  \quad
\includegraphics[width=7.6cm]{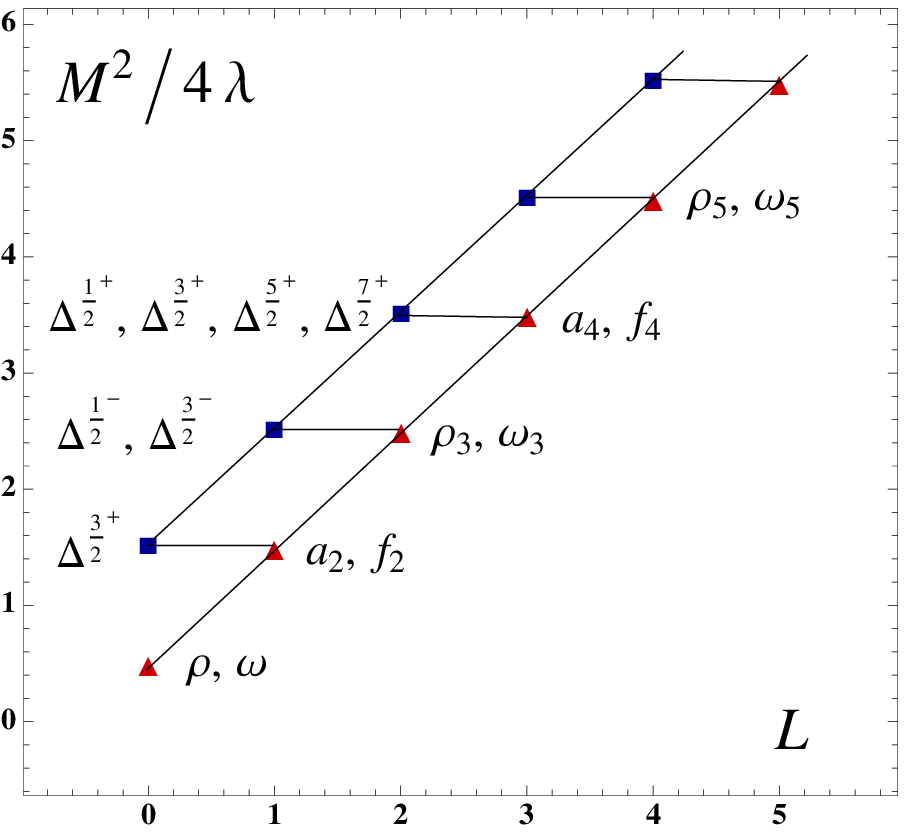}
\end{center}
\caption{\label{MNSCC} \small Meson-baryon superconformal connection.  The predicted values of $M^2$ in units of $4 \la$ for pseudoscalar mesons (red triangles) and nucleons  (blue squares) are plotted vs the orbital angular momentum $L$.  The operator $R_\la$ acts horizontally to the right and transforms a baryon with orbital $L$ into a meson with orbital $L + 1$. $R_\la^\dagger$ acts on the opposite direction and annihilates the $\pi$ meson which has no nucleon partner (left).  The $\rho$ and $\omega$ have no baryonic partner, since it would imply a negative value of $L$ (right). The hadron quantum number assignment is taken from Ref.~\cite{Brodsky:2014yha}.}
\end{figure}
%%%%%%%%%%%%%%%%%%

The fermion operator  $R^\dagger$ transforms a meson state $\vert M, L  \rangle$ with orbital angular momentum $L$ into a nucleon state $ \vert N, L - 1\rangle$ with orbital $L-1$: The superpartners have the same parity. The operator $R^\dagger$ also annihilates the lowest possible mass state, the pion, which has no nucleon superpartner ~\cite{Dosch:2015nwa}:
\beq
R^\dagger \vert M, L  \rangle = \vert N, L - 1\rangle, \quad \quad R^\dagger \vert M, L = 0 \rangle = 0.
\enq
Thus the special role of  the pion in the supersymmetric approach to hadronic physics as a unique state of zero energy (See Fig. \ref{MNSCC} (left)).

\section{Baryon-meson dynamical supersymmetry \label{BMS}}

For light hadronic systems the essential confinement dynamics is encoded in the supercharge generator $R_\la$ which connects meson and nucleon bound-state wavefunctions.   We expect a rather similar situation for the baryon superpartner trajectory of the  $\rho$ and $\omega$ vector mesons, the $\Delta$ baryons, the essential difference being an overall shift of the spectrum due to the spin interaction (See Fig. \ref{MNSCC} (right)). This is indeed the case experimentally, where the superconformal predictions are reproduced to a very good degree of accuracy~\cite{Dosch:2015nwa}.  A crucial prediction observed in Fig. \ref{MNSCC} is the fact that the lowest meson state in each family with $L = 0$ has no superpartner. Physically, this corresponds to the fact that the partner baryon would have $L = -1$. Indeed, the lowest meson state is annihilated by the supercharge $R_\la^\dagger$ and there is no baryon partner~\footnote{A few questions left open in Ref.~\cite{Dosch:2015nwa}, notably the nature of the spin interaction and the correct twist assignment of the wave functions in the $\rho-\Delta$ connections, have been addressed recently by enforcing superconformal symmetry in the holographic embedding of the corresponding light-front bound-state equations~\cite{BdTDL}.}.

It is remarkable that the underlying dynamical supersymmetry holds across the heavy-light hadron spectrum~\cite{Dosch:2015bca}, even if conformal symmetry is badly broken by heavy quark masses.   In this case the superpotential  is no longer  constrained by conformal symmetry and it is basically unknown. However, the supersymmetric predictions for mesons and baryons with charm and beauty are also observed to a surprising level of accuracy~\cite{Dosch:2015bca}. In particular, the crucial prediction for the lowest state $L = 0$ meson, namely that it has no supersymmetric baryon partner --as it would correspond to an $L = -1$ state, is confirmed in the observed heavy-light hadron spectrum~\cite{Dosch:2015bca}.

\section{Concluding remarks}

The classical QCD Lagrangian in the limit of zero-quark mass  has no scale.  The origin of a mass scale and confinement remains essentially a fundamental unsolved problem  due to the vast complexity of the strong dynamics of quarks and gluons in the low energy domain~\cite{Brambilla:2014jmp, Greensite:2011zz}. Indeed the problem is of such complexity that it may become~ undecidable~\cite{Cubitt:2015nat}.  

The connection of light-front dynamics, classical gravity  in a higher-di\-men\-sional space and superconformal quantum mechanics brings new insights to our understanding of  confinement dynamics and QCD.  In particular, an indication of the emergence of a mass scale in a conformal invariant theory can be drawn from the mechanism proposed by de Alfaro, Fubini and Furlan~\cite{deAlfaro:1976je} in the context of one-dimensional QFT.  Mathematically, this remarkable result follows from the isomorphism of the group $SO(2,1)$ of conformal QM with the one-dimensional conformal group $Conf\left(R^1\right)$. Since the generators of  $Conf\left(R^1\right)$, $H$, $D$ and $K$, have different dimensions, their relation with the generators of $SO(2,1)$ requires the introduction of a scale $\sqrt \la$. The scale $\sqrt \la$ plays the role of the confinement scale, which is directly related to a physical observable, such as a hadron mass.   In this context, scales corresponding to scheme dependent perturbative computations at short distances, such as the QCD renormalization scale $\La_s$, are determined by the nonperturbative scale $\sqrt \la$~\cite{Deur:2014qfa}.  In superconformal QM the Hamiltonian $G$ \req{G} is the superposition of generators with different dimensions which also requires the introduction of a scale $\sqrt \la$~\cite{Fubini:1984hf}.

The mapping of the superconformal Hamiltonian to the light-front relativistic bound-state equations u\-nique\-ly determines the confinement potential $U$, including constant terms in the potential. It also determines uniquely the dilaton profile $\varphi$ in the embedding space.  The holographic connection leads to the extension of the LF potential $U$ to arbitrary spin~\cite{deTeramond:2013it}. This embedding is required to describe consistently all the light mesons and baryons and their superconformal connections~\cite{BdTDL}. For heavy quark masses conformal symmetry is broken but the supersymmetric predictions are observed all across the heavy-light spectrum~\cite{Dosch:2015bca}, giving strong evidence of an underlying dynamical supersymmetry in hadron physics.

There is another important connection which we have not described here: the connection between the one-dimensional conformal group $Conf\left(R^1\right)$ and the group of isometries of AdS$_2$ in two dimensions, $SO(2,1)$, which underlies the AdS$_2$/CFT$_1$ correspondence~\cite{Chamon:2011xk}.   AdS$_2$ becomes relevant to the emergence of its one-dimensional dual QFT in the infrared through holographic renormalization in the bulk~\cite{Faulkner:2009wj}. Indeed the holographic flow of boundary theories to AdS$_2$ geometry in the infrared~\cite{Heemskerk:2010hk} may be relevant to the approach of hadron physics discussed in this article,  since its dual one-dimensional quantum field theory is realized as conformal quantum mechanics.

\section*{Acknowledgments}

The results presented here are based on collaborations with  Stanley J. Brodsky, Alexandre Deur, Hans Guenter Dosch and C\'edric Lorc\'e. I
want to thank the organizers of Light Cone 2015 for their hospitality at  Frascati.

\end{document}